\documentclass[showpacs,preprintnumbers,amsmath,amssymb]{revtex4}
\usepackage{graphicx}
\usepackage{txfonts}
\usepackage{epsfig}
\usepackage{graphicx}
\usepackage{dcolumn}
\usepackage{bm}
\preprint{APS/123-QED}
\begin{document}
\title{{\large The Length of the Day: {\it A Cosmological
Perspective}}}
\author{Arbab I. Arbab}
\email{aiarbab@uofk.edu} \affiliation{Department of Physics,
Faculty of Science, University of Khartoum, P.O. 321, Khartoum
11115, SUDAN} \affiliation{Department of Physics and Applied
Mathematics, Faculty of Applied Sciences and Computer, Omdurman
Ahlia University, P.O. Box 786, Omdurman, Sudan}
\date{\today}
\begin{abstract}
We have found an empirical law for the variation of the length of
the Earth's day with geologic time employing Wells's data. We
attribute the lengthening of the Earth's day to the present cosmic
expansion of the Universe. The prediction of law has been found to
be in agreement with the astronomical and geological data. The day
increases at a present rate of 0.002 sec/century. The length of the
day is found to be 6 hours when the Earth formed. We have also found
a new limit for the value of the Hubble constant and the age of the
Universe.
\end{abstract}
 \maketitle
\section{Introduction}
\label{sec:intro} According to Mach's principle the inertia of an
object is not a mere property of the object but depends on how
much matter around the object. This means that the distant
universe would affect this property. Owing to this, we would
expect a slight change in the strength of gravity with time. This
change should affect the Earth-Moon-Sun motion. It is found that
the length of the day and the number of days in the year do not
remain constant. From coral fossil data approximately 400 million
years (m.y.) ago, it has been  estimated that there were little
over 400 days in a year at that time. It is also observed that the
Moon shows an anomalous acceleration (Dickey, 1994). As the
universe expands more and more matter appears in the horizon. The
expansion of the universe may thus have an impact on the
Earth-Moon-Sun motion. Very recently, the universe is found to be
accelerating at the present time (Peebles, 1999, Bahcall {\it et
al.}, 1999). To account for this scientists suggested several
models. One way to circumvent this is to allow the strength of
gravity to vary slightly with time (Arbab, 2003). For a flat
universe, where the expansion force is balanced by gravitational
attraction force, this would require the universe to accelerate in
order to avoid a future collapse. This can be realized if the
strength of the gravitational attraction increases with time
(Arbab, 1997, 2003), at least during the present epoch (matter
dominated). One appropriate secure way to do this is to define an
effective Newton's constant, which embodies this variation while
keeping the `bare' Newton's constant unchanged. The idea of having
an effective constant, which shows up when a system is interacting
with the outside world, is not new. For instance, an electron in a
solid moves not with its `bare' mass but rather with an effective
mass. This effective mass exhibits the nature of interaction in
question. With the same token, one would expect a celestial object
to interact (couple) with its effective constant rather than the
normal Newton's constant, which describes the strength of gravity
in a universe with constant mass. We, therefore, see that the
expansion of the universe affects indirectly (through Newton's
constant) the evolution of the Earth-Sun system. Writing an
effective quantity is equivalent to having summed all
perturbations (gravitational) affecting the system. With this
minimal change of the ordinary Newton's constant to an effective
one, one finds that Kepler's laws can be equally applicable to a
perturbed or an unperturbed system provided the necessary changes
are made.  Thus one gets a rather modified Newton's law of
gravitation and Kepler's laws defined with this effective constant
while retaining their usual forms. In the present study, we have
shown that the deceleration of the Earth rotation is, if not all,
mainly a cosmological effect. The tidal effects of the Earth
deceleration could, in principle,  be a possible consequence, but
the cosmological consequences should be taken seriously.
\\
The entire history of the Earth has not been discovered so far.
Very minute data are available owing to difficulties in deriving
it. Geologists  derived some information about the length of the
day in the pats from the biological growth rhythm preserved in the
fossil records (e.g., bi-valves, corals, stromatolites, etc). The
first study of this type was made by the American scientist John
Wells (1963), who investigated the variation of the number of days
in the year from the study of fossil corals. He inferred, from the
sedimentation layers of calcite  made by the coral, the number of
days in the year up to the Cambrian era. Due to the lack of a
well-preserved records, the information about the entire past is
severely hindered. The other way to discover the past rotation is
to extrapolate the presently observed one. This method, however,
could be very misleading.
\section{The Model}
\label{sec:using} Recently, we proposed a cosmological model for
an effective Newton's constant (Arbab, 1997) of the form
\begin{equation}G_{\rm eff.}=G_0\left(\frac{t}{t_0}\right)^\beta\ ,
\end{equation}
where the subscript `0' denotes the present value of the quantity:
$G_0$ is the normal (bare) Newton's constant and $t_0$ is the
present age of the Universe. Here $G_{\rm eff.}$ includes all
perturbative effects arising from all gravitational sources. We
remark here that $G_0$ does not vary with time, but other
perturbations induce an effect that is parameterized in $G_{\rm
eff.}$ in the equation of motion. Thus, we don't challenge here
any variation in the normal Newton's constant $G_0$. We claim that
such a variation can not be directly measured as recently
emphasized by Robin Booth (2002). It can only be inferred from
such analysis. We remark here that $\beta$ is not well determined
($\beta
> 0$) by the cosmological model. And since the dynamics of the
Earth is determined by Newton's law of gravitation any change in
$G$ would affect it. This change may manifest its self in various
ways. The length of day may attributed to geological effects which
are in essence gravitational. The gravitational interaction should
be described by Einstein's equations. We thus provide here the
dynamical reasons for these geological changes. We calculate the
total effect of expansion of the universe on the Earth dynamics.

The Kepler's 2nd law of motion for the Earth-Sun system,
neglecting the orbit eccentricity, can be written as
\begin{equation}
G_{\rm eff.}^2[(M+m)^2m^3]\ T_{\rm eff.} = 2\pi L_{\rm eff.}^3\ \
,
\end{equation}
where $m$, $M$ are the mass of the Earth and the Sun respectively;
$L_{\rm eff.}$ is the orbital angular momentum of the Earth and
$T_{\rm eff.}$ is the period (year) of the Earth around the Sun at
any time in the past measured by the days in that time. $T_{\rm
eff.}$ defines the number of days (measured at a given time) in a
year at the epoch in which it is measured. This is because the
length of day is not constant but depends on the epoch in which it
is measured. Since the angular momentum of the Earth about the Sun
hasn't changed, the length of the year does not change. We however
measure the length of the year by the number of days which are not
fixed. The length of the year in seconds (atomic time) is fixed.
Thus one can still use Kepler's law as in eq.(2)(which generalizes
Kepler's laws) instead of adding other perturbations from the
nearby bodies to the equation of motion of the Earth. We, however,
incorporate all these perturbations in a single term, viz. $G_{\rm
eff.}$. Part of the total effect of the increase of length of day
could show up in geological terms. We calculate here the total
values affecting the Earth dynamics without knowing exactly how
much the contribution of each individual components.

The orbital angular momentum of the Earth (around the Sun) is
nearly constant.
 From equation (2), one can write
\begin{equation}
T_{\rm eff.}=T_0\left(\frac{G_0}{G_{\rm eff.}}\right)^2\ ,
\end{equation}
where $T_0=365$ days and $G_0=6.67\times 10^{-11}\rm\ N m^2
kg^{-2}$.
\\
Eqs.(1) and (3) can be written as
\begin{equation}\label{1}
T_{\rm eff.}=T_0\left( \frac{t_0-t_p}{t_0}\right)^{2\beta}\ ,
\end{equation}
where $t_0$ is the age of the universe and $t_p$ is the time
measured from present time backward. This equation can be casted
in the form
\begin{equation}\label{1}
x=\ln\left(\frac{T_{\rm eff.}}{T_0}\right)=2\beta\ln\left(
\frac{t_0-t_p}{t_0}\right) ,
\end{equation}
or equivalently,
\begin{equation}
t_0=\frac{t_p}{\left(1-\exp(-x/2\beta)\right)} ,
\end{equation}
To reproduce the data obtained by Wells for the number of days in
a year (see Table.1), one would require $\beta=1.3$ and
$t_0\backsimeq 11\times 10^9$ years. This is evident since, from
(Arbab, 2003) one finds the Hubble constant is related to the age
of the Universe by the relation,

\begin{equation}\label{4}
    t_0=\left(\frac{2+\beta}{3}\right)H_0^{-1}=1.1\  H_0^{-1}\ ,
\end{equation}
and the effective Newton's constant would vary as
\begin{equation}
G_{\rm eff}=G_0\left(\frac{t_0-t_p}{t_0}\right)^{1.3}.
\end{equation}
This is an interesting relation, and it is the first time relation
that constrained the age of the Universe (or Hubble constant)from
the Earth rotation. However, the recent Hipparcos satellite
results (Chaboey {\it et al} 1998) indicate that the age of the
universe is very close to 11 billion years. Hence, this work
represent an unprecedented confirmation for the age of the
universe. One may attribute that the Earth decelerated rotation is
mainly (if not only) due to cosmic expansion that shows up in
tidal deceleration. Thus, this law could open a new channel for
providing valuable information about the expansion of the
Universe. The Hubble constant in this study amounts to $H_0=\rm
97.9\ km\ s^{-1} Mpc^{-1} $. However, the Hubble constant is
considered to lie in the limit, $\rm 50\ \ km\ s^{-1} Mpc^{-1}
<H_0< \rm 100\ km\ s^{-1} Mpc^{-1} $. Higher values of $H_0$ imply
a fewer normal matter, and hence a lesser dark matter.  This
study, therefore, provides an unprecedented way of determining the
Hubble constant. Astronomers usually search into the space to
collect their data about the Universe. This well determined value
of $\beta$ is crucial to the predictions of our cosmological model
in Arbab, 2003. We notice that the gravitational constant is
doubled since the Earth was formed (4.5 billion years ago).
\\
From eqs.(3) and (8) one finds the effective number of days in the
year ($T_{\rm eff.}$) to be
\begin{equation} T_{\rm
eff.}=T_0\left(\frac{t_0}{t_0-t_p}\right)^{2.6}\ ,
\end{equation}
and since the length of the year is constant, the effective length
of the day ($D_{\rm eff.}$) is given by
\begin{equation}
\ D_{\rm eff.}=D_0\left(\frac{t_0-t_p}{t_0}\right)^{2.6}\ ,
\end{equation}
so that
\begin{equation}
T_0D_0=T_{\rm eff.}D_{\rm eff.} \ \ \ .
\end{equation}
We see that the variation of the length of day and month is a
manifestation of the changing conditions (perturbation) of the
Earth which are parameterized as a function of time ($t)$ only.
Thus, equation (7) guarantees that the length of the year remains
invariant.
\section{Discussion} The Wells's fossil data is shown in
Table 1 and our corresponding values are shown in Table 2. In
fact, the length of the year does not change, but the length of
the day was shorter than now in the past. So, when the year is
measured in terms of days it seems as if the length of the year
varies. Sonett {\it et al.} (1996) have shown that the length of
the day 900 m.y ago was 19.2 hours, and the year contained 456
days. Our law gives the same result (see Table 2). Relying on the
\emph{law of spin isochronism} Alfv\'{e}n and Arrhenius (1976)
infer for the primitive Earth a length of day of 6 hours (p.226).
Using coral as a clock, Poropudas (1991, 1996) obtained an
approximate ancient time formula based on fossil data. His formula
shows that the number of days in the year is 1009.77 some 3.556
b.y. ago. Our law shows that this value corresponds rather to a
time 3.56 b.y. ago, and that the day was 8.7 hours. He suggested
that the day to be 5 - 7 hours at approximately 4.5 b.y. ago.
Ksanfomality (1997) has shown that according to the principle of
isochronism all planets had an initial period of rotation between
6 - 8 hours. However, our model gives a value of 6 hours (see
Table 2). Berry and Baker (1968) have suggested that laminae,
ridges and troughs, and bands on present day and Cretaceous
bivalve shells are growth increments of the day and month,
respectively. By counting the number of ridges and troughs they
therefore find that the  year contains 370.3 days in the late
Cretaceous. According to the latest research by a group of Chinese
scientists (Zhu \emph{et al}.), there were 15 hours in one day,
more than 540 days, in a year from a study of \emph{stromatolite}
samples. We however remark that according to our law that when the
day was 15 hours there were 583 days in a year 1.819 billion years
ago. The difference in time could be due to dating of their rock.
\\
Recently, McNamara and Awramik (1992) have concluded, from the
study of {\it stromatolite}, that at about 700 m.y. ago the number
of days in a year was $\rm 435$\ days and the length of the day
was $\rm 20.1$\ hours. In fact, our model shows that this value
corresponds more accurately to $\rm 715$\ m.y. ago. Vanyo and
Awramik (1985) has investigated {\it stromatolite}, that is 850
m.y. old, obtained a value between 409 and 485 days in that year.
Our law gives 450 days in  that year and 19.5 hours in that day.
This is a big success for our law. Here we have gone over all data
up to the time when the Earth formed. We should remark that this
is the first model that gives the value of the length of the day
for the entire geologic past time.
\\
The present rate of increase in the length of the day is $0.002 \
\rm m s/ century$. Extrapolating this astronomically determined
lengthening of the day since the seventeenth century leads to 371
days in the late Cretaceous (65 m.y. ago). The slowing down in the
rotation is not uniform; a number of irregularities have been
found. This conversion of Earth's rotational energy into heat by
tidal friction will continue indefinitely making the length of the
day longer. In the remote past the Earth must have been rotating
very fast. As the Earth rotational velocity changes, the Earth
will adjust its self to maintain an equilibrium (shape) compatible
with the new situation. In doing so, the Earth should have
experienced several geologic activities. Accordingly, one would
expect that the tectonic movements (plate's motion) to be
attributed to this continued adjustment.
\\
We plot the length of day (in hours) against time (million years
back) in Fig.(1). We notice here that a direct extrapolation of
the present deceleration would bring the age of the Earth-Moon
system t a value of 3.3 billion years. We observe that the plot
deviates very much from straight line. The plot curves at two
points which I attribute the first one to emergence of water in
huge volume resulting in slowing down the rotation of the Earth's
spin. The second point is when water becomes abundant and its rate
of increase becomes steady. These two points correspond to\ 1100
m.a. and 3460 m.a., and their corresponding lengths of day are
18.3 and 8.9 hours, respectively. As the origin of life is
intimately related to existence of water, we may conclude that
life has started since 3.4 billion years ago, as previously
anticipated by scientists.
\section{Conclusion}
We have constructed a model for the variation of length of the day
with time. It is based on the idea of an effective Newton's
constant as an effective coupling representing all gravitational
effects on a body. This variation can be traced back over the
whole history of the Earth. We obtained an empirical law for the
variation of the length of the day and the number of days in a
year valid for the entire past Earth's rotation. We have found
that the day was 6 hours when the Earth formed. These data
pertaining to the early rotation of the Earth can  help
paleontologists to check their data with this findings. The change
in the strength of gravity is manifested in the way it influences
the growth of biological systems. Some biological systems
(rythmites, tidalites, etc) adjust their rhythms with the lunar
motion (or the tide). Thus any change in the latter system will
show up in the former. These data can be inverted and used as a
geological calendar. The data we have obtained for the length of
the day and the number of days in the year should be tested
against any possible data pertaining to the past's Earth rotation.
Our empirical law has been tested over an interval as far back as
4500 m.y. and is found to be in consistency with the experimental
data so far known. In this work we have arrived at a generalized
Kepler's laws that can be applicable to our ever changing
Earth-Moon-Sun system.
\section*{Acknowledgements}
I wish to thank the  University of Khartoum  for providing research
support for this work, and the Abdus salam International Center for
Theoretical Physics (ICTP) for hospitality where this work is
carried out.
\section{References}
\hspace{-0.4cm}
Alfv\'{e}n, H and Arrhenius, G., 1976. \emph{Evolution of the solar system}, NASA, Washington, USA \\
Arbab, A.I., 1997. {\it Gen. Relativ. Gravit. 29}, 61.\\
Arbab, A.I., 2003. {\it Class. Quantum. Gravit. 20}, 93.\\
Bahcall, N.A., {\it et al.} 1999. {\it Science 284}, 1481.\\
Berry, W.B. and Barker, R.M., 1968. {\it Nature 217}, 938.\\
Chaboyer, B. {\it et al.}, 1998. {\it Astrophys. J., 494},96. \\
Dickey, J.O., et al., 1994. {\it Science 265}, 482.\\
Ksanfomality, L.V., 1997. {\it Astrophys.Space Sci. 252}, 41.\\
McNamara, K.J, Awramik, S.M., 1992. {\it Sci. Progress 76}, 345.
\\
Pannella, G., 1972. {\it Astrophys. Space Sci. 16}, 212.\\
Peebles, J., 1999. {\it Nature 398}, 25.\\
Poropudas, H. K. J., 1996. {\it Harrastelijan ajatuksia päivän,
kuukauden ja vuoden pituudesta muinaisina aikoina. Geologi, 4-5},
92.
\\
Poropudas, H. K. J., 1991.
http://www.cs.colorado.edu/~lindsay/creation/coral-clocks.txt.
\\
Booth Robin, http://arXiv.org/abs/gr-qc/0203065\\
Sonett, C.P., 1996. Kvale, E.P., Chan, M.A. and Demko,T.M., {\it Science, 273}, 100.\\
Vanyo, J. P. and Awramik, S. M., 1985. {\it Precambrian Research, 29}, 121.\\
Wells, J.W., 1963. {\it Nature, 197}, 948.\\
Zhu, S.X.  Huang, X.G.  and Xin, H.T., 2002.  \emph{Proceedings of
the 80th Anniversary of the Chinese Geological Society}, Geology
Publishing House, Beijing.
\newpage
\begin{table}
\caption{Data obtained  from  fossil corals and radiometric time
(Wells, 1963) } \vspace{1cm}
\begin{tabular}{|r|r|r|r|r|r|r|r|r|r|}
\hline
Time$^*$  & 65 & 136 & 180 & 230 & 280 & 345 & 405 & 500 & 600\\
\hline
solar days/year & 371.0 & 377.0 & 381.0 & 385.0 & 390.0 & 396.0 & 402.0 & 412. 0 & 424.0\\
\hline
\end{tabular}
\end{table}
\begin{table}
\caption{Data obtained from our empirical law: eqs.(9) and (10)}
\vspace{1cm}
\begin{tabular}{|r|r|r|r|r|r|r|r|r|r|}
\hline
Time$^*$   & 65 & 136 &  180  & 230 & 280  & 345  & 405 & 500 & 600 \\
\hline
solar days/year  & 370.9 & 377.2 & 381.2 & 385.9 & 390.6 & 396.8 & 402.6 & 412.2  & 422.6\\
\hline
length of solar day (hr) & 23.6 & 23.2 & 23.0 & 22.7 & 22.4 & 22.1 & 21.7 & 21.3 & 20.7\\
\hline
\end{tabular}
\end{table}
\begin{table}
\begin{tabular}{|r|r|r|r|r|r|r|r|r|r|}
\hline
Time$^*$   & 715 & 850 & 900  & 1200  & 2000 & 2500 & 3000 & 3560 & 4500\\
\hline
solar days/year  & 435.0 & 450.2 & 456  & 493.2  & 615.4 & 714.0 & 835.9 & 1009.5 & 1434\\
\hline
length of solar day (hr) & 20.1  & 19.5 & 19.2  & 17.7  & 14.2& 12.3 & 10.5 & 8.7 & 6.1\\
\hline
\end{tabular}
\vspace{1cm}\\  $^*$ Time is measured in million years (m.y.)
before present.
\end{table}
\label{lastpage}
\newpage
\includegraphics[width=6.9in]{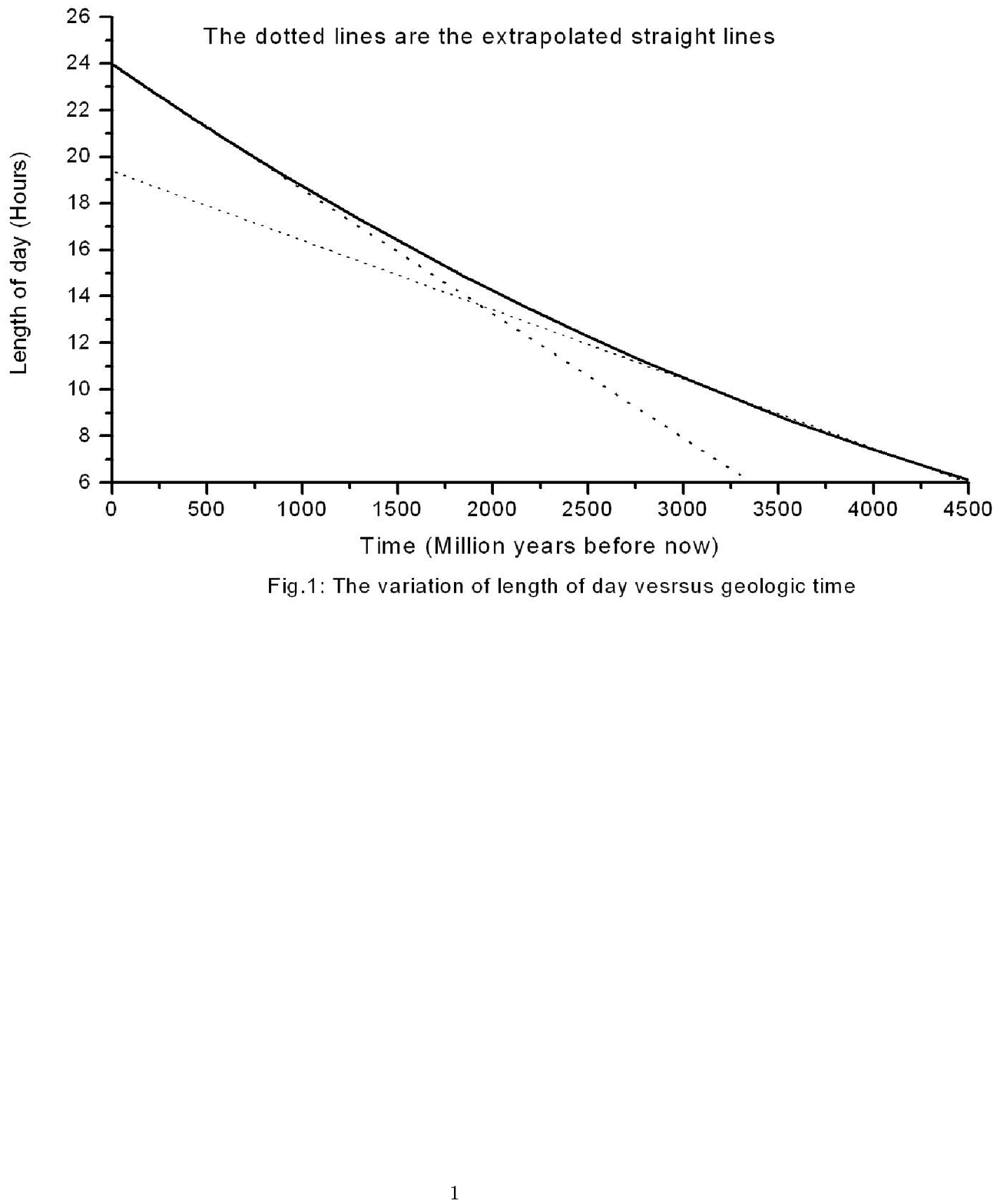}
\end{document}